\documentclass[reprint,amsmath,amssymb]{revtex4-2}
\usepackage[utf8]{inputenc}

\usepackage{bm}
\usepackage{graphicx}
\usepackage{upgreek}
\usepackage{amsmath}
\usepackage{amssymb}
\usepackage{numprint}
\usepackage{xspace}
\usepackage{comment}
\usepackage[dvipsnames]{xcolor}

\definecolor{myred}{rgb}{0.78, 0, 0}
\definecolor{myblue}{rgb}{0, 0, 0.78}

\usepackage{hyperref}
\hypersetup{colorlinks = true, citecolor = myblue, linkcolor = myblue}

\newcommand{\diff}{\mathrm{d}}
\newcommand{\vect}{\boldsymbol}
\newcommand{\tens}{\boldsymbol}

\newcommand{\mean}[1]{\langle #1 \rangle}
\newcommand{\abs}[1]{|#1|}
\newcommand{\Eqref}[1]{Eq.~\ref{#1}}
\newcommand{\figref}[1]{Fig.~\ref{#1}}

\DeclareMathOperator{\arctanh}{arctanh}
\DeclareMathOperator{\cov}{cov}
\DeclareMathOperator{\tr}{tr}
\newcommand{\jac}{{\tens J}(\vect q)}
\newcommand{\Tjac}{\tilde{\tens J}(\vect q)}
\newcommand{\qMax}{\vect q_\mathrm{max}}

\newcommand{\nocontentsline}[3]{}
\newcommand{\tocless}[2]{\bgroup\let\addcontentsline=\nocontentsline#1{#2}\egroup}

\begin{document}

\title{Physical interactions promote Turing patterns}

\author{Lucas Menou%
  \email{lucas.menou@ds.mpg.de}}
\thanks{L.M. and C.L. contributed equally to this work.}

\author{Chengjie Luo%
  \email{chengjie.luo@ds.mpg.de}}
\thanks{L.M. and C.L. contributed equally to this work.}

\author{David Zwicker%
  \email{david.zwicker@ds.mpg.de}}
\affiliation{Max Planck Institute for Dynamics and Self-Organization, Am Faßberg 17,
  37077 Göttingen, Germany}

\date{\today}

\begin{abstract}
Turing's mechanism is often invoked to explain periodic patterns in nature, although direct experimental support is scarce. Turing patterns form in reaction-diffusion systems when the activating species diffuse much slower than the inhibiting species, and the involved reactions are highly non-linear. Such reactions can originate from co-operativity, whose physical interactions should also affect diffusion. We here take direct interactions into account and show that they strongly affect Turing patterns. We find that weak repulsion between the activator and inhibitor can substantially lower the required differential diffusivity and reaction non-linearity. In contrast, strong interactions can induce phase separation, but the resulting length scale is still typically governed by the fundamental reaction-diffusion length scale. Taken together, our theory connects traditional Turing patterns with chemically active phase separation, thus describing a wider range of systems. Moreover, we demonstrate that even weak interactions affect patterns substantially, so they should be incorporated when modeling realistic systems.
\end{abstract}

\maketitle

\tableofcontents

\section{Introduction}
Natural periodic patterns, ranging from nano-crystals~\cite{Fuseya2021}, tissues~\cite{Recho2019}, populations dynamics~\cite{Karig2018}, to geophysical phenomena~\cite{Goehring2013}, are often explained by the seminal Turing mechanism~\cite{Turing1952,Vittadello2021,Kondo2010,Cross2009}.
Turing patterns generally describe the spatial distribution of an activator and an inhibitor that diffuse in space. %
Patterns then form when the localized activator triggers production while the inhibitor suppresses production globally, often summarized as \emph{local activation, global inhibition}~\cite{Vittadello2021, Kondo2010}.
However, it is not clear whether Turing's mechanism can actually explain natural patterns~\cite{Kondo2022, Maini2012} since inhibitors need to diffuse much faster than activators and the involved reactions need to be highly non-linear~\cite{Haas2021, Diambra2015, Gierer1972}.
Such non-linear reactions are often motivated by co-operative reactions, where multiple reactants lead to non-linearities~\cite{Diambra2015}.
Co-operativity typically originates from physical interactions, which should also affect the diffusive motion of the species, but this is typically not taken into account.

Physical interactions are crucial for organizing biomolecules in cells~\cite{Banani2017}, cells in tissues~\cite{Tsai2022}, and even organisms in groups~\cite{Liu2013}.
In particular, multivalent interactions can induce phase separation, where a dense droplet phase segregates spontaneously from a dilute surrounding phase.
This is possible since the enthalpic gain from the interactions overcompensates the entropic loss of concentrating constituents~\cite{Choi2020a,Dignon2020}.
In simple passive systems, surface tension implies that large droplets grow to system size at the expense of small droplets~\cite{Voorhees1992}.
However, chemical reaction can suppress this Ostwald ripening~\cite{Zwicker2015} and thus control the size and arrangement of droplets \cite{Soeding2019,kirschbaum2021control,Hondele2020}.
The predicted hexagonal arrangements~\cite{Motoyama1996} are very reminiscent of Turing patterns, although patterns are driven by phase separation in these systems.
Taken together, although droplets regulated by chemical reactions share some properties with Turing patterns, it is unclear how the two models are related. 

In this paper, we study a minimal system that is capable of forming droplets as well as Turing patterns.
Effectively, we add physical interactions between activator and inhibitor to a standard reaction-diffusion system.
This approach allows us to quantify how interactions affect pattern formation and it unveils a range of systems that interpolate between Turing's mechanism and patterns formed in active phase separating systems.
In particular, we show how weak repulsive interactions stabilize patterns by inducing cross-diffusion, while strong interactions lead to phase separation, where coarsening is arrested by chemical reactions.

\section{Results}

\subsection{Interactions affect pattern formation}

We start by considering a minimal pattern forming system comprised of two species: an activator~$A$ and an inhibitor~$I$.
The basic Turing model describes the dynamics of the respective fractions~$\phi_A(\vect r, t)$ and $\phi_I(\vect r, t)$ as a function of the spatial position~$\vect r$ and time $t$,
\begin{align}
	\label{eqn:pde_turing}
	\partial_t \phi_i & =
		\sum_{j=A,I} \mathcal D_{ij} \nabla^2 \phi_j
		+ k\left[ \frac{2 \phi_0}{1 + (\frac{\phi_I}{\phi_A})^h} - \phi_i\right]
\end{align}
for $i=A,I$.
Here, the first term on the right hand side describes ideal diffusion with a diffusivity matrix $\mathcal D_{ij}$ and the second term captures chemical reactions based on the Hill–Langmuir equation~\cite{Weiss1997}.
For $h\ge 1$, these reactions promote the production of $A$ and $I$ by activator $A$ and suppress it by the inhibitor~$I$, while both species exhibit linear degradation with rate~$k$.
This choice of chemical reactions allows us to independently control the typical fraction~$\phi_0$ of components $A$ and $I$, the reaction rate~$k$, and the reaction non-linearity~$h$.
We show in the Supporting Information that \Eqref{eqn:pde_turing} exhibits a Turing instability for sufficiently large~$h$ and diffusivity ratio~$D_I/D_A$, so the inhibitor spreads out while the activiator stays localized.

\begin{figure*}
  \centering
  \includegraphics[width=\textwidth]{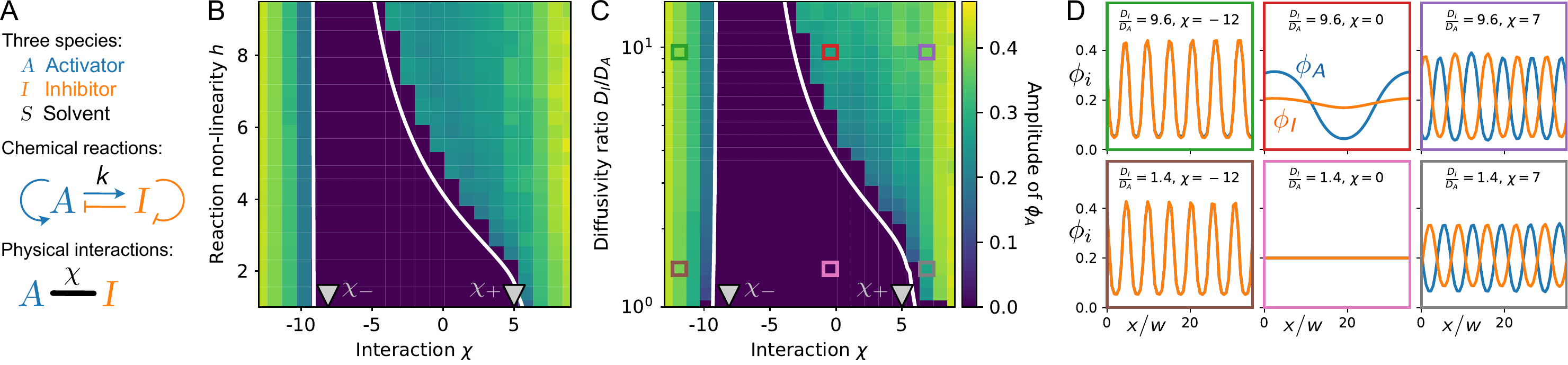}
  \caption{\textbf{Interactions affect pattern formation.}
  (A) Schematic of chemical and physical interactions of activator~$A$, inhibitor~$I$, and the inert solvent.
  (B) Stationary state amplitudes of fraction~$\phi_A$ as a function of the interaction strength~$\chi$ and the reaction non-linearity~$h$ for diffusivity ratio $D_I / D_A = 5$.
  (C) Amplitude as a function of~$\chi$ and $D_I/D_A$ for $h=5$.
  (D)~Stationary patterns of $\phi_A$ (blue) and $\phi_I$ (orange) for the indicated parameters.
  (B, C)
  The homogeneous state is stable between the white lines, obtained from a linear stability analysis of \Eqref{eqn:pde}, and the gray triangles mark critical interaction values $\chi_-$ and $\chi_+$; see Supporting Information.
  (B--D)
  Model parameters are $k=0.1\, D_A/w^2$ and $\phi_0 = 0.2$.
  Simulations ran for $t=10^5 w^2/D_A$ on a 1D grid of length $200 \, w$ with periodic boundary conditions. %
  } 
  \label{fig:overview} 
\end{figure*}

To include physical interactions between activator $A$ and inhibitor $I$, we first consider the thermodynamics of 
an incompressible, isothermal fluid comprising the species~$A$ and $I$ as well as an inert solvent~$S$; see \figref{fig:overview}A.
This  system is still fully described by the volume fractions $\phi_A$ and $\phi_I$, since the solvent occupies the remaining fraction~$\phi_S = 1 - \phi_A - \phi_I$.
The interactions of $A$ and $I$ in such a fluid can then be described by the Flory-Huggins free energy~\cite{safran2018statistical, rubinstein2003polymer, Cahn1958}
\begin{align}
	\label{eqn:free_energy}
	F[\phi_A, \phi_I] &= \frac{k_\mathrm{B} T}{\nu} \int\Bigl[
		\phi_A \ln\phi_A + \phi_I \ln\phi_I + \phi_S \ln\phi_S
\notag\\&\quad
		+\chi \phi_A \phi_I
		+ \frac{w^2}{2} \left( |\nabla \phi_A|^2 + |\nabla \phi_I|^2\right)
	\Bigr]\diff \vect r
	\;,
\end{align}
where the integral is over the volume of the system, $k_\mathrm{B} T$ is the relevant energy scale, and $\nu$ denotes a molecular volume, which we assume to be the same for all species.
The first three terms in the square bracket capture the translational entropies of all species, the fourth term describe the physical interaction between $A$ and $I$, and the last term limits the width of interfaces between coexisting phases to roughly $w$ in strongly interacting systems~\cite{Cahn1958}.
The interactions between $A$ and $I$ are quantified by the Flory parameter~$\chi$:
Positive~$\chi$ denotes effective repulsion, which can originate from heterotypic repulsion or homotypic attraction, while negative~$\chi$ leads to attraction of $A$ and $I$.
Equilibrium states, which minimize the free energy given by \eqref{eqn:free_energy}, can be inhomogeneous when interactions are sufficiently strong~\cite{weber2019physics}:
For strong attraction (large negative $\chi$), a phase enriched in $A$ and $I$ will segregate from one enriched in the solvent, whereas strong repulsion (large positive $\chi$) will lead to segregation of $A$ from $I$ with an equal amount of solvent in both phases.
However, it is unclear how this equilibrium behavior is modified by the active reactions described by \eqref{eqn:pde_turing} and how Turing patterns are affected by weak interactions.

To model reaction-diffusion dynamics with interactions described by the Flory parameter~$\chi$, we replace the ideal diffusion term in \Eqref{eqn:pde_turing} by a more general form which describes diffusion in non-ideal fluids.
Linear non-equilibrium thermodynamics implies that diffusive fluxes are then proportional to gradients of the chemical potentials associated with the free energy given by \Eqref{eqn:free_energy}, and the proportionality constants (known as Onsager coefficients or mobilities) determine the kinetic rate~\cite{Zwicker2022a,julicher2018hydrodynamic}.
Defining  non-dimensional exchange chemical potentials~$\mu_i = \nu (k_\mathrm{B} T)^{-1} \delta F/\delta \phi_i$,
\begin{subequations}
\label{eqn:chemical_potentials}
\begin{align}
    \mu_A &= \ln(\phi_A) - \ln(\phi_S) + \chi \phi_I - w^2 \nabla^2 \phi_A
\\
    \mu_I &= \ln(\phi_I) - \ln(\phi_S) + \chi \phi_A - w^2 \nabla^2 \phi_I
    \;,
\end{align}
\end{subequations}
we thus find
\begin{align}
	\label{eqn:pde}
	\partial_t \phi_i & =
		\nabla . (D_i \phi_i \nabla \mu_i)
		+ k\left[ \frac{2 \phi_0}{1 + (\frac{\phi_I}{\phi_A})^h} - \phi_i\right]
		\;,
\end{align}
where $D_i$ are the diffusivities of the species $i=A,I$, which are related to the mobilities $D_i\phi_i$ in this multicomponent system~\cite{Kramer1984}.
We show in the Supporting Information that \Eqref{eqn:pde} reduces to \Eqref{eqn:pde_turing}, and thus describes ideal diffusion, if physical interactions are absent ($\chi=0$) and the wave length of patterns is large compared to $w$.
Consequently, \eqref{eqn:pde} describes a reaction-diffusion system encompassing non-ideal diffusion and containing normal Turing patterns as a limiting case.

To see how interactions affect patterns, we performed numerical simulations of Eqs.~(\ref{eqn:chemical_potentials})--(\ref{eqn:pde}) in a one-dimensional system with periodic boundary conditions; see Methods.
\figref{fig:overview} demonstrates that without interactions ($\chi=0$), patterns with finite amplitudes only emerge if the reactions are sufficiently non-linear (large $h$) and the inhibitor diffuses sufficiently fast ($D_I \gg D_A$), as expected for Turing patterns~\cite{Turing1952}.
This trend persists for weak interactions, although the corresponding threshold values of $h$ and $D_I/D_A$ change.
Apparently, repulsion between $A$ and $I$ promotes pattern formation ($\chi  > 0$), while attraction suppresses it ($\chi < 0$).
However, very strong attraction can again lead to large amplitudes ($\chi \lesssim -9$), independent of $h$ and $D_I/D_A$. %

The corresponding volume fraction profiles shown in \figref{fig:overview}D corroborate these observations:
Without interactions (middle column), the system stays either homogeneous (pink parameter set) or forms normal Turing patterns (red parameter set) with a localized activator~$A$ and a fairly homogeneous inhibitor~$I$.
In contrast, strong attraction (left column) leads to co-localization of $A$ and $I$, reminiscent of phase separation, albeit with a well defined pattern length scale.
Similarly, strong repulsion (right column) implies segregation of $A$ from $I$.
Taken together, we thus showed that there is an interesting interplay between stereotypical Turing patterns and interactions promoting phase separation.
 
\subsection{Weak interactions imply cross-diffusion}

To understand how interactions affect pattern formation, we first analyze weak interactions ($\abs{\chi} \lesssim 5$) by treating them as a perturbation to normal Turing patterns. %
Assuming the wave length of patterns is large compared to $w$, the generalized diffusion in \Eqref{eqn:pde} can be approximated by ideal diffusion to first order in $\chi$; see Supporting Information.
Consequently, the dynamics are described by \Eqref{eqn:pde_turing} with the diffusivity matrix
\begin{align}
	\label{eqn:diffusivity_matrix}
	\mathcal D_{ij} &\approx 
\left(
\begin{array}{cc}
	D_A (1+ \psi) &
	D_A (\psi  + \chi\phi_0 ) \\[5pt]
	D_I (\psi + \chi\phi_0 ) &
	D_I (1+ \psi) \\
\end{array}
\right) \;,
\end{align} 
where $\psi = \phi_0/(1 - 2 \phi_0)$.
This analysis demonstrates that without interactions ($\chi=0$) in dilute system ($\phi_0\ll1$, avoiding crowding effects), diffusion is dominated by the diagonal entries, resulting in stereotypical Turing patterns.
In this case, we show analytically in the Supporting Information that patterns can only form when the diffusivity ratio $D_I/D_A$ and the reaction non-linearity $h$ are sufficiently large, consistent with the literature~\cite{Gierer1972,Diambra2015}.

\begin{figure}
  \centering
  \includegraphics[width=\columnwidth]{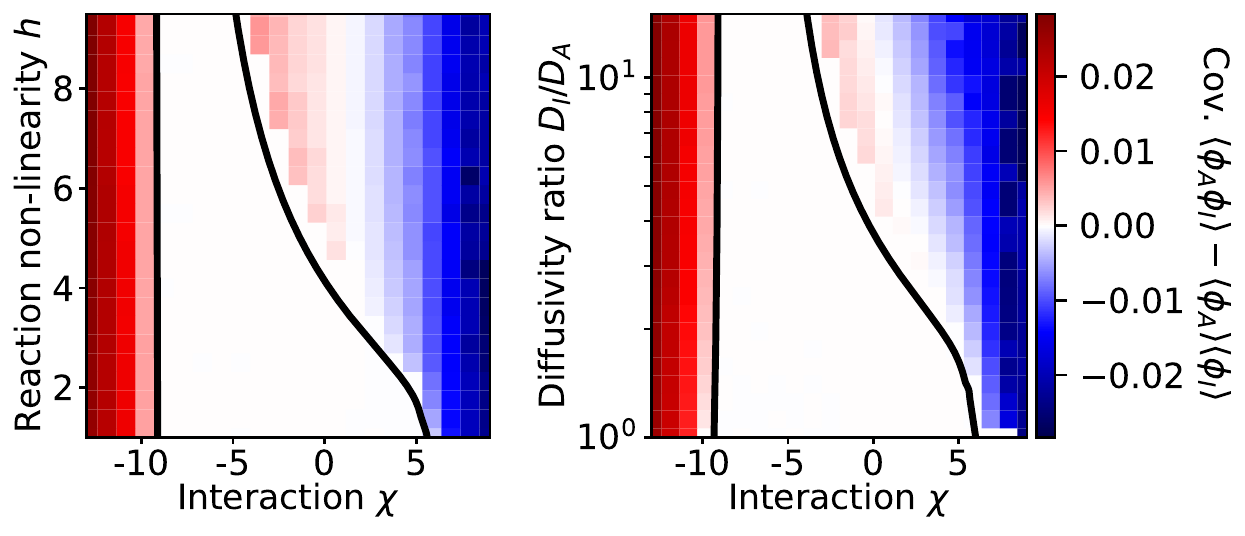}
  \caption{\textbf{Interactions affect correlation between activator and inhibitor.}
  Covariance $\mean{\phi_A\phi_I} - \mean{\phi_A}\mean{\phi_I}$  as a function of interaction~$\chi$ and reaction non-linearity $h$ (left, $D_I/D_A=5$) or diffusivity ratio $D_I/D_A$ (right, $h=5$).
    The homogeneous state is stable between the black lines, obtained from a linear stability analysis of \Eqref{eqn:pde}; see Supporting Information.
    Model parameters are   $k=0.1\, D_A/w^2$ and $\phi_0 = 0.2$.
    Simulations ran for $t=10^5 w^2/D_A$ on a periodic 1D grid of length $200 \, w$.
} 
  \label{fig:correlations}
\end{figure}

\begin{figure}
  \centering
  \includegraphics[width=0.64\columnwidth]{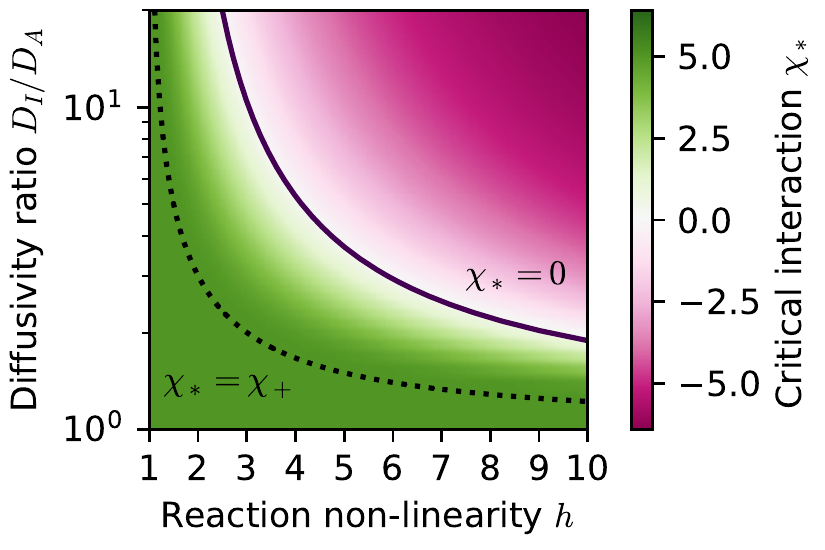}
  \caption{\textbf{Repulsive interactions improve trade-off between differential diffusivity and reaction non-linearity.}
    Minimal interaction~$\chi_*$ to support patterns (Eq. S11 in the Supporting Information) as a function of diffusivity ratio~$D_I/D_A$ and reaction non-linearity~$h$ for $\phi_0=0.2$.
    Turing patterns without interactions ($\chi_*=0$) form above the solid line.
    Conversely, phase separation is required to form patterns below the dotted line ($\chi_* > \chi_+=5$).
} 
  \label{fig:critical_chi}
\end{figure}

If $A$ and $I$ interact ($\chi \neq 0$), \Eqref{eqn:diffusivity_matrix} reveals that interactions directly affect cross-diffusion of $A$ and $I$.
For example, repulsive interactions ($\chi>0$) imply fluxes of $A$ opposite to the gradient of $I$, thus favoring the segregation of the two species and enhancing patterns~\cite{Vanag2009}.
To quantify this behavior, we analyze the covariance, $\cov(\phi_A, \phi_I)  = \mean{\phi_A\phi_I} - \mean{\phi_A}\mean{\phi_I}$, where the brackets denote spatial averages in the stationary state.
\figref{fig:correlations} shows that the covariance generally decreases with more repulsive interactions (larger $\chi$, consistent with enhanced cross-diffusion), increasing reaction non-linearity $h$, and diffusivity ratio~$D_I/D_A$.
The more detailed stability analysis presented in the Supporting Information demonstrates that repulsive interactions always promote pattern formation and lower the required reaction non-linearity~$h$ and diffusivity ratio $D_I/D_A$; see \figref{fig:critical_chi}.
In contrast, attractive interactions ($\chi < 0$) generally stabilize the homogeneous system. %
However, this behavior only holds for moderate interactions~$\chi$ since strong attraction ($\chi \lesssim -9$) also leads to large amplitudes; see Fig.~\ref{fig:overview}.

\subsection{Strong interactions invoke phase separation}
Turing's mechanism cannot explain patterns that form when the activator $A$ and inhibitor~$I$ attract each other strongly ($\chi < -9$).
Based on the strong correlations between $A$ and $I$ seen in \figref{fig:correlations}, we hypothesize that strong attraction leads to associative phase separation of $A$ and $I$ from the solvent, while the reactions play a minor role.
Indeed, we show in the Supporting Information that phase separation is possible in the absence of reactions when $\chi < \chi_-$, where $\chi_- = 
8 \arctanh(1-4 \phi_0 )/(4 \phi_0 -1)$ marks the binodal point for a given average fraction $\phi_0$ of $A$ and $I$.
\figref{fig:overview} shows that the value $\chi_-$ is very close to the onset of patterns, and that the resulting profiles are  perfectly co-localized.
Taken together, strong attraction between $A$ and $I$ leads to co-segregation of the two components from the solvent while the reactions barely affect the amplitude.

Strong repulsion between $A$ and $I$ should also lead to phase separation.
In fact, for $\chi > \chi_+$ with $\chi_+ = 1/\phi_0$, we predict that $A$ can segregate from $I$ spontaneously, even without reactions present; see Supporting Information.
Figs.~\ref{fig:overview}--\ref{fig:correlations} suggest that increasing the repulsion beyond this points indeed results in strong anti-correlation between $A$ and $I$ and a vanishing threshold for $h$ and $D_I/D_A$.
This numerical data indicates a continuous transition from patterns formed by reactions and diffusion (Turing patterns for weak interactions) to those formed by phase separation (strong interactions, $\chi < \chi_-$ or $\chi > \chi_+$).
Taken together, we showed that patterns can form by reactions and by phase separation with an intricate interplay between them.

\subsection{Reaction rate controls pattern length scale}

\begin{figure*}
  \centering
  \includegraphics[width=\textwidth]{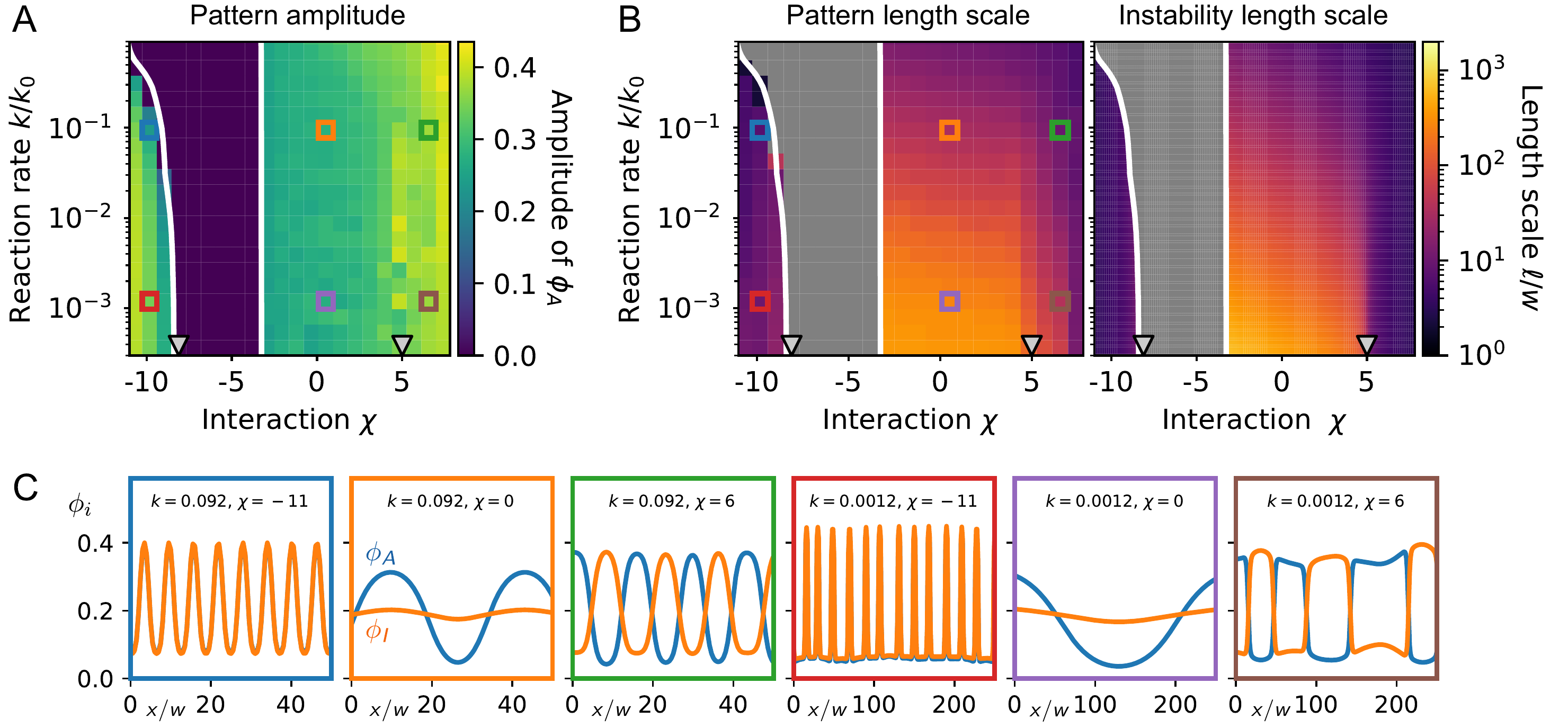}
  \caption{\textbf{Reaction rate~$k$ determines pattern length scale}.
  (A) Amplitude of activator~$\phi_A$ as a function of interaction~$\chi$ and reaction rate~$k$.
  (B) Pattern length scale~$\ell$ determined from the maximum of the structure factor of $\phi_A$ from numerical simulations (left) and from the fast growing mode in a linear stability analysis (right) as a function of $\chi$ and $k$
  (C)~Stationary patterns of $\phi_A$ (blue) and $\phi_I$ (orange) for various parameters indicated in panels A and B.
  (B, C)
  The homogeneous state is stable between the white lines, obtained from a linear stability analysis of \Eqref{eqn:pde}, and the gray triangles mark critical interaction values $\chi_-$ and $\chi_+$; see Supporting Information.
  (A--C) Model parameters are $h = 5$, $D_I/D_A=10$, $\phi_0 = 0.2$, and $k_0 = D_A/w^2$.
  Simulations ran for $t=10^7 k_0^{-1}$ on a periodic 1D grid of length $2000 \, w$.
}
  \label{fig:reaction}
\end{figure*}

We next ask what determines the length scale~$\ell$ of the patterns.
We show in the Supporting Information that $\ell$ is hardly affected by variations of the reaction non-linearity $h$ and diffusivity ratio $D_I/D_A$. %
In contrast, the interaction strength~$\chi$ has a stronger influence: More repulsive interactions lead to patterns with shorter wave lengths, presumably because larger $\chi$ promote pattern formation. 
However, the strongest influence on the pattern length scale~$\ell$ is the reaction rate~$k$:
Numerical simulations and the linear stability analysis presented in \figref{fig:reaction} indicated that $k$ allows adjusting $\ell$ over several orders of magnitude with barely any changes in the pattern amplitude.

To understand how the reaction rate~$k$ affects the pattern length scale~$\ell$, we first focus on weak interactions.
In this case, interactions mainly cause cross-diffusion (see \Eqref{eqn:diffusivity_matrix} and Supporting Information), implying that the reaction diffusion lengths $\sqrt{D_A/k}$ and $\sqrt{D_I/k}$ are the only length scales in the equations.
Consequently, length scales in the stationary state and in the initial instability must scale with $k^{-1/2}$ for weak interactions, consistent with \figref{fig:reaction}B.
For strong interactions ($\chi<\chi_-$ or $\chi > \chi_+$), the system exhibits phase separation, implying that the initial instability is dominated by short patterns of length~$w$ while the stationary state patterns may exhibit much longer length scales due to coarsening~\cite{Bray1994}.
\figref{fig:reaction}B shows that the linear stability analysis indeed predicts $\ell \sim w$ in the region where we predict phase separation.
For associative phase separation at strong attraction ($\chi < \chi_-$) these patterns remain stable and coarsening is suppressed; the variation in $\ell$ reflects the influence of $\chi$ on the interfacial width; see Supporting Information.
In the contrasting case of strong repulsion ($\chi > \chi_+$), patterns coarsen to the reaction-diffusion length and thus scale with $k^{-1/2}$; see \figref{fig:reaction}B.
This behavior is similar to the coarsening observed in active droplets, where the final length scale is also governed by the reaction-diffusion length~\cite{Zwicker2015}.
Note that the numerical data presented in \figref{fig:reaction}B might not represent the full stationary state since domain sizes only grow logarithmically with time in these 1D systems~\cite{Bray1994}.
However, our analysis demonstrates that the length scale~$\ell$ of the final pattern is generally governed by reaction-diffusion lengths, except when strong attraction between $A$ and $I$ leads to associative phase separation.

\subsection{Results generalize to higher dimensions}
\begin{figure*}
  \centering
  \includegraphics[width=2\columnwidth]{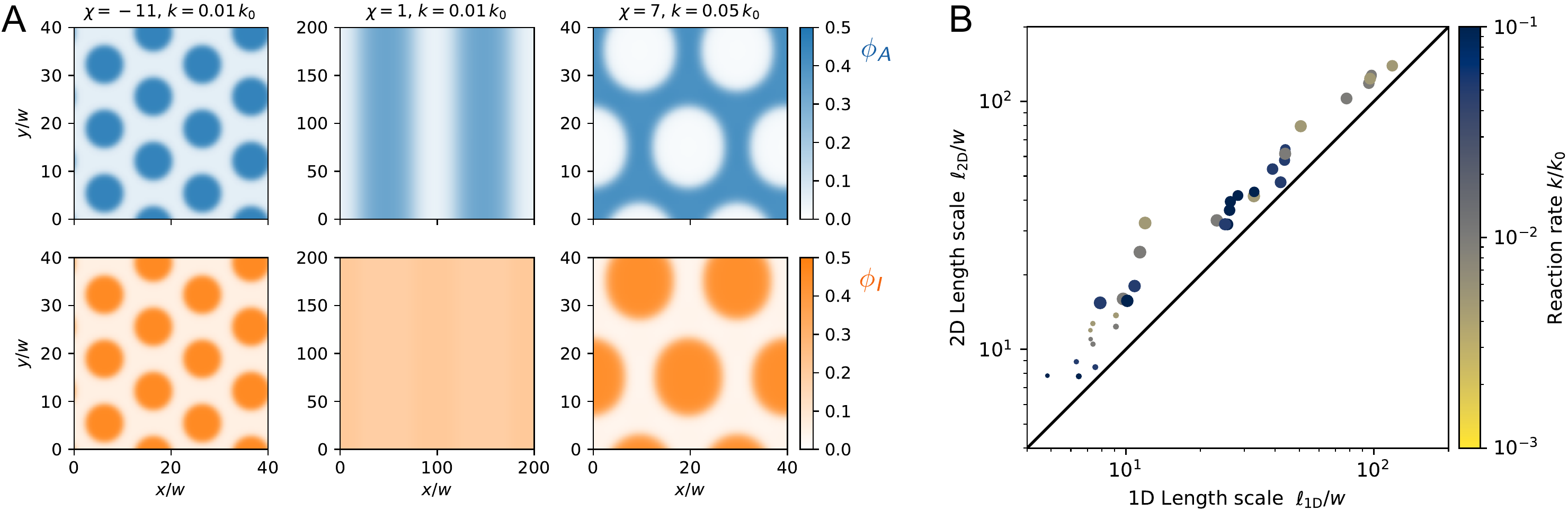}
  \caption{\textbf{Interactions also control patterns in higher dimensions.}
  (A) 2D stationary patterns of $\phi_A$ (upper panels) and $\phi_I$ (lower panels) for strong attraction ($\chi=-11$), weak interaction ($\chi=1$), and  strong repulsion ($\chi=7$) from left to right.
  (B) Correlation of length scales~$\ell$ determined from numerical simulations in 1D and 2D for various reaction rates~$k$ (color scale) and interactions~$\chi$ (marker size).
  (A--B) Additional model parameters are $h = 5$, $D_I/D_A=10$, $\phi_0 = 0.2$, and $k_0 = D_A/w^2$.
} 
  \label{fig:2d_pattern}
\end{figure*}

So far, we have focused on pattern formation in one dimension for simplicity, but many natural patterns form in planar geometries.
To see whether our results hold for this relevant case, we next perform a few selected simulations in two dimensions; see \figref{fig:2d_pattern}.
Analogously to one dimension, we find Turing patterns for weak interactions (middle column of \figref{fig:2d_pattern}A) and strong interactions induce phase separation.
In particular, strong attraction between $A$ and $I$ leads to co-localization (left column) whereas strong repulsion induces anti-correlated patterns (right column).
Interestingly, in both cases of phase separation droplets form instead of stripe patterns, even though both phases occupy roughly half of the space.
In such a case, normal phase separating systems exhibit stripe patterns~\cite{Matsen2012}, but the reaction-diffusion dynamics in our system apparently alter the picture. 
In any case, \figref{fig:2d_pattern}B shows that the pattern length scales we measured in 1D are very close to the ones measured in 2D simulations, implying that the results from the simple 1D system translate to the more complex 2D system.

\section{Discussion}
We propose an extension to Turing patterns that take into account physical interactions that occur naturally.
Weak repulsion between activator and inhibitor enhances patterns by inducing cross-diffusion, thus amplifying local activation and global inhibition.
In contrast, strong interactions lead to phase separation, which can either be associative ($A$ and $I$ co-localize) or segregative ($A$ separates from $I$).
Both cases exhibit patterns for a much larger range of diffusivities and reaction non-linearities than normal Turing patterns, and the resulting length scales differ:
In the segregative case of strong repulsion, patterns are governed by the reaction-diffusion length scale and thus grow larger for weaker reactions.
In contrast, patterns in the associative case of strong attraction are arrested at the interfacial width.
Taken together, we thus demonstrated that interactions can affect patterns substantially.
The linear stability analysis presented in the Supporting Information demonstrates that these results do not depend on the specific choice of the reactions.
Instead, interactions can generally lift restrictions on diffusivities and reaction non-linearities imposed by ordinary Turing patterns.
Since physical interactions are virtually always present, many natural patterns can likely be explained by similar mechanisms.

Physical interactions in natural systems can stem from various sources and are virtually unavoidable in multicomponent systems.
We need to investigate such  systems in more detail, both in terms of physical interactions~\cite{Zwicker2022a}, chemical reaction networks~\cite{Haas2021,Diego2018}, and conservation laws~\cite{Brauns2020}.
For instance, Turing patterns can form when two species have equal diffusivity, while a third one is immobile~\cite{Marcon2016}, to produce effective differences in diffusivities.
Explaining natural patterns in detail also requires incorporating growth~\cite{Werner2015}, flows~\cite{bhattacharyya2021coupling}, noise, and delays~\cite{Maini2012}.
Moreover, natural patterns often form in complex geometries, including coupled layers~\cite{Yang2003} and curved surfaces~\cite{Nishide2022}, where the mechano-chemical coupling~\cite{Veerman2021} can lead to dynamic patterns~\cite{Krause2021}.
The organization of biological cells is a particularly exciting example since biomolecules are known to interact and react~\cite{Banani2017}.
While this sometimes leads to spatial patterns explained by Turing's mechanism~\cite{Kondo2010} other examples are akin to active droplets~\cite{Soeding2019}.
Another possibility is patterns formed by self-propelled agents, which can exhibit motility-induced phase separation~\cite{Cates2015a}, and explain some population patterns successfully~\cite{Liu2013}.
In all these cases, physical interactions will affect patterns qualitatively and quantitatively, opening new perspectives on how natural patterns emerge.

\tocless{\subsection{Methods} \label{subsec:methods}}

We perform numerical simulations of \Eqref{eqn:pde} on an equidistantly discretized grid using second-order finite-differences to approximate differential operators~\cite{Zwicker2020}.
We evaluate $\nabla\mu$ on a staggered grid to ensure material conservation and use an explicit Euler scheme for the time evolution.

\tocless{\subsection{Acknowledgement}}
We thank Pierre Haas, Riccardo Rossetto, Noah Ziethen, and Yicheng Qiang for helpful discussions and critical reading of the manuscript.
We gratefully acknowledge funding from the Max Planck Society and the European Union (ERC, EmulSim, 101044662).

\begin{appendix}
\label{appendix}
\renewcommand{\thefigure}{S\arabic{figure}}
\setcounter{figure}{0}

\section{Linear stability analysis of full model}
\label{sec:linear_stability}

\begin{figure}
  \centering
  \includegraphics[width=\columnwidth]{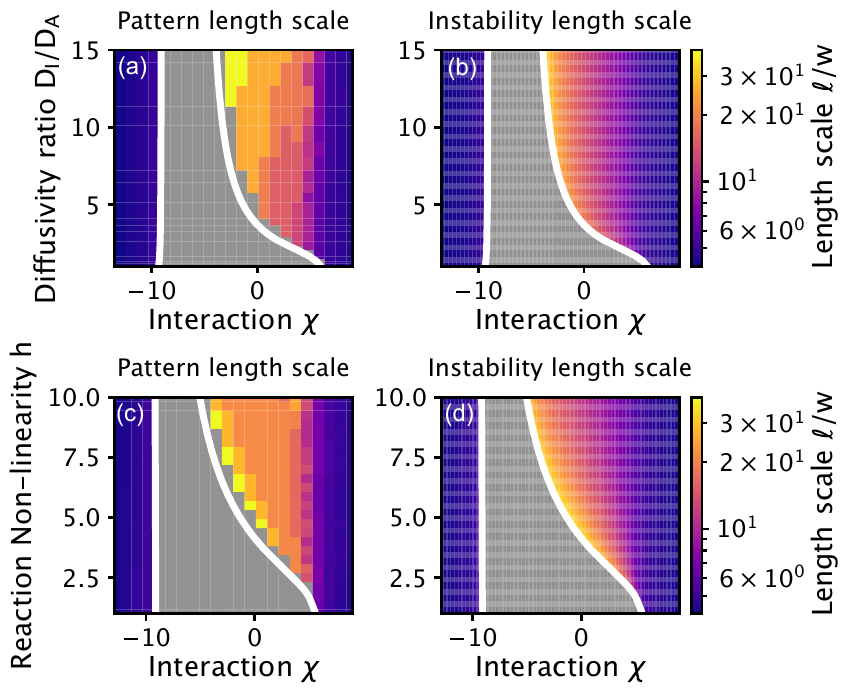}
  \caption{
  Pattern length scale~$\ell$ determined from the maximum of the structure factor of $\phi_A$ determined from numerical simulations (left) and predicted by a linear stability analysis (right).
  The upper panels show $\ell$ as a function of interaction $\chi$ and diffusivity ratio $D_I/D_A$ for reaction non-linearity $h=5$.
  The lower panels show $\ell$ as a function of $\chi$ and $h$ for $D_I/D_A=10$.
  Remaining model parameters are $k=0.1\, D_A/w^2$ and $\phi_0 = 0.2$.
  Note that length scales do not vary much as a function of $h$ and $D_I/D_A$, while $\chi$ has a significant influence.
  } 
  \label{fig:overview_length_scale} 
\end{figure}

We here present details of the linear stability analysis of the dynamical equations, given by \Eqref{eqn:pde} in the main text, for the volume fractions for the activator, $\phi_A(\vect r, t)$, and  the inhibitor, $\phi_I(\vect r, t)$.
The homogeneous stationary state is given by $\phi_A=\phi_I=\phi_0.$ 
We linearize \Eqref{eqn:pde} in the main text around the uniform solution and determine the time evolution of perturbations in Fourier space, where the perturbations are characterized by the wave vector~$\vect q$.
Their stability is then determined by the eigenvalues of the Jacobian matrix~\cite{Cross2009}
\begin{widetext}
 \begin{equation}
  \label{eq:S_matrix_stab}
	\jac = \left(
	\begin{array}{cc}
		-\vect q^2 D_A\left(1+\psi+\phi _0 w^2\vect q^2 \right)+\frac{(h-2) k}{2} &
		-\vect q^2 D_A \left(\psi+\phi _0 \chi \right)-\frac{h k}{2} \\[5pt]
		 -\vect q^2  D_I \left(\psi+\phi _0\chi \right)+\frac{h k}{2} &
		 -\vect q^2 D_I \left(1+\psi+\phi _0 w^2\vect q^2 \right)-\frac{(h+2) k}{2} \\
	\end{array}
\right) \;, 
\end{equation}
\end{widetext}
where we defined $\psi = \phi_0/(1 - 2 \phi_0)$, which is the ratio of the fraction of species $A$ and $I$ to the fraction of the solvent.
The two eigenvalues of $\jac$ are given by
 \begin{eqnarray}
 \label{eq:S_eigens}
	\sigma^{\pm}(\vect q) = \frac{\tr\jac}{2} \pm\frac{1}{2}\sqrt{(\tr\jac )^2-4 \det\jac}
	\;.
 \end{eqnarray}
 For an instability to occur, the real part of one of the eigenvalues must be positve.
The real part of $\sigma^-(\vect q)$ is always negative since $\tr \jac =-\left[\vect q^2 (D_A+D_I)\left(1+\psi+\phi _0 w^2 \vect q^2\right)+2k\right]<0$.
In contrast, the real part of $\sigma^+(\vect q)$ is positive when $\det\jac < 0$.
To homogeneous state is thus unstable if there is at least one value of $\vect q$ for which this is the case and the bifurcation lines in Fig. \ref{fig:overview}--\ref{fig:reaction} in the main text can be determined numerically solving the condition
 \begin{equation}
  \min_{\vect q} \det\jac = 0
  \;.
  \label{eq:S_stab_eq}
\end{equation}
Moreover, we can determine the length scale $\ell=2\pi/q_{\mathrm{max}}$ associated with the most unstable mode, where the rate $\sigma^+(\vect q_\mathrm{max})$ is maximal.
\Eqref{eq:S_eigens} shows that the maximum of $\jac$ coincides with the minimum of $\det\jac$, allowing us to determine $\ell$ using numerical minimization.
Although the linear stability analysis only provides information at the initial stage when the volume fractions are close to $\phi_0$, the final length scale of the pattern will often be similar to the instability length scale; see \figref{fig:reaction} in the main text and \figref{fig:overview_length_scale}.

\section{Limit of weak interactions}
\label{sec:weak_interactions}
To compare our extended model with physical interactions (\Eqref{eqn:pde} of the main text) to ordinary Turing systems (\Eqref{eqn:pde_turing} of the main text), we show a formal reduction.

\subsection{Expansion of full model for weak interactions}
We thus consider the limit of weak perturbations, where the entropic contributions of the solvent (third term in the integrand in \Eqref{eqn:free_energy} of the main text) and the gradient terms (last term in the integrand) are negligible.
In this case, we find that the diffusive terms in \Eqref{eqn:pde} reduce to ideal diffusion with the diffusivity matrix
\begin{align}
  \tilde{\mathcal D}
=
  \left(
\begin{array}{cc}
  D_A\left(1+\psi\right)&  D_A\left( \psi+\phi_0\chi\right) \\
 D_I\left( \psi+\phi_0\chi\right) & D_I\left(1+\psi\right)\\
\end{array}
\right)
\;,
\end{align}
where $\psi = \phi_0/(1 - 2 \phi_0)$.
The dynamics in this limit are described by \Eqref{eqn:pde_turing} in the main text with $\tilde{\mathcal D}$ as the diffusivity matrix.
In particular, the stability of the homogeneous state $\phi_A=\phi_I = \phi_0$ follows from the eigenvalues of the matrix
 \begin{align}
  \label{eq:S_matrix_stab}
\tilde{  \vect {J}}(\vect q)\! =\! \left(\!\!\!
\begin{array}{cc}
-q^2 D_A\left(1+\psi\right)+\frac{(h-2) k}{2}& -q^2 D_A \left(\psi+\phi _0 \chi \right)-\frac{h k}{2} \\
 -q^2  D_I \left(\psi+\phi _0\chi \right)+\frac{h k}{2} & -q^2 D_I \left(1+\psi \right)-\frac{(h+2) k}{2} \\
\end{array}\!\!
\right) 
, 
 \end{align}
 which is the same as $ \vect {J}(\vect q)$ given in \Eqref{eq:S_matrix_stab} except for the ${\vect q}^4$ terms.
The bifurcation lines can again be determined using \Eqref{eq:S_stab_eq}.
\figref{fig:h_chi_h_DI} shows that the resulting bifurcation lines are very similar to those determined for the full model, particularly for $\chi \approx 0$, thus validating the approximation.

\subsection{Linear stability analysis of approximate model}
We next determine bifurcation lines by solving \Eqref{eq:S_stab_eq}.
The determinant of $\jac$ is a quadratic polynomial in $\vect q^2$,
\begin{align}
	\det(\Tjac) &= a (\vect q^2)^2 + b \vect q^2 +c
\end{align}
with
\begin{subequations}
\begin{align}
	a &= \frac{\eta  D_A^2 \left(\chi  \phi _0^2 \left(\chi -2 \chi  \phi _0+2\right)-1\right)}{2 \phi _0-1}
\\
	b &= -\frac{k D_A}{2 - 4 \phi_0} \Bigl[
		2(\eta +1)(\phi_0-1) 
\notag\\&\qquad+ (\eta -1)(\phi _0 h \chi (1- 2 \phi _0)+ h)
	\Bigr]
\\
	c &= k^2
	\;,
\end{align}
\end{subequations}
where $\eta = D_I/D_A$ is the diffusivity ratio.
To find the minimum of $\det(\Tjac)$, we can distinguish two cases based on the sign of $a$.
For $a<0$, the minimum of $\det(\Tjac)$ is at $\vect q^2=+ \infty$ and always negative.
The condition $a<0$ reduces to $\chi <\chi^-_\mathrm{spin}$ or $\chi >\chi^+_\mathrm{spin}$, where
\begin{align}
	\chi^-_\mathrm{spin} &=-\frac{1}{ \phi _0(1-2\phi _0)}
	&\text{and} &&
	\chi^+_\mathrm{spin}=\frac{1}{\phi _0}
	\;,
\end{align}
consistent with the analysis of the spinodal lines presented in Section~\ref{sec:passive_system}.

\begin{figure}[b]
  \centering
  \includegraphics[width=\columnwidth]{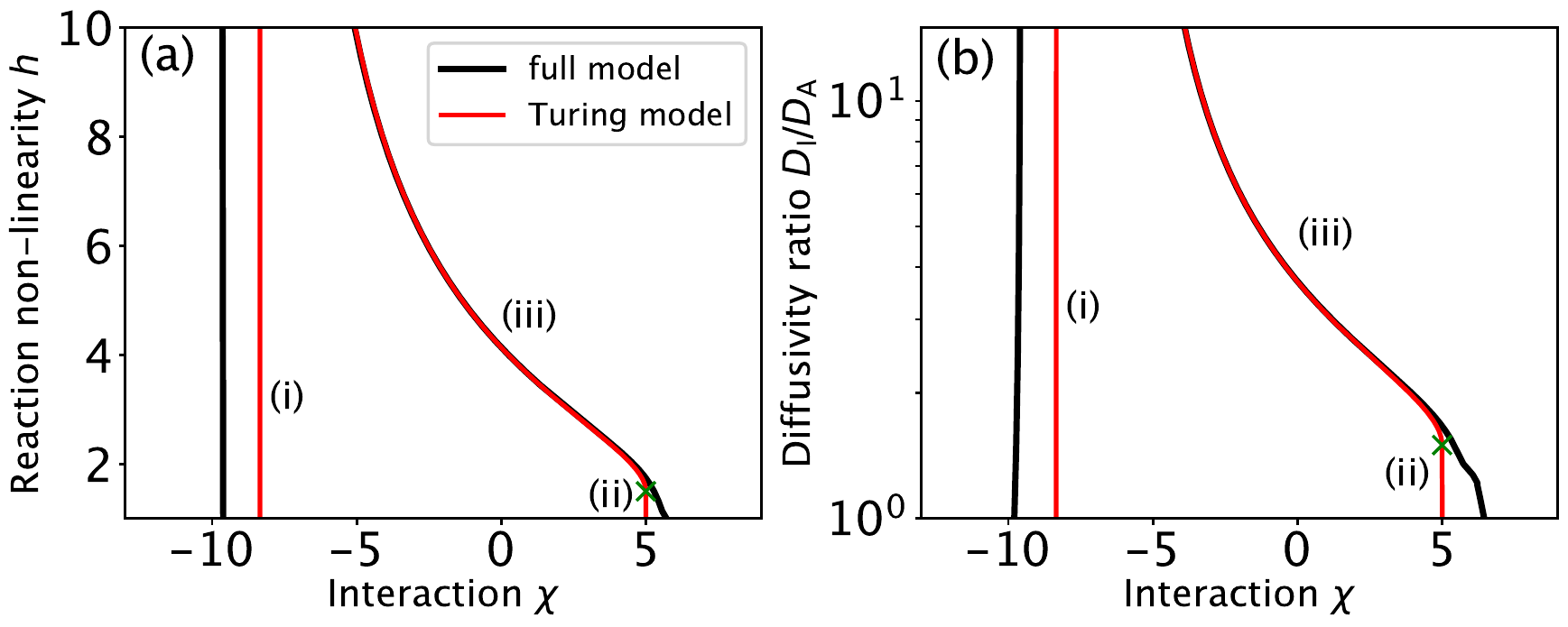}
  \caption{Bifurcation lines of the full model (black curves) and the effective Turing model (red curves). The red lines in each panel constitute of three parts, (i) $\chi=-1/\left(\phi_0(1-2\phi_0)\right)\approx-8.33$, (ii) $\chi=1/\phi_0=5$, and (iii) data calculated using Eq.~\ref{eq:S_chi_sol}. The green cross is to indicate the point where (ii) and (iii) meet, which is at $h=(\eta+1)/(\eta-1)=1.5$ in panel (a) and $\eta=D_\mathrm{I}/D_\mathrm{A}=(h+1)/(h-1)=1.5$ in panel (b). Other model parameters are the same as in Fig.~1(b) for panel (a) and in Fig.~1(c) for panel (b). } 
  \label{fig:h_chi_h_DI} 
\end{figure}

\begin{figure}[b]
  \centering
  \includegraphics[width=0.8\columnwidth]{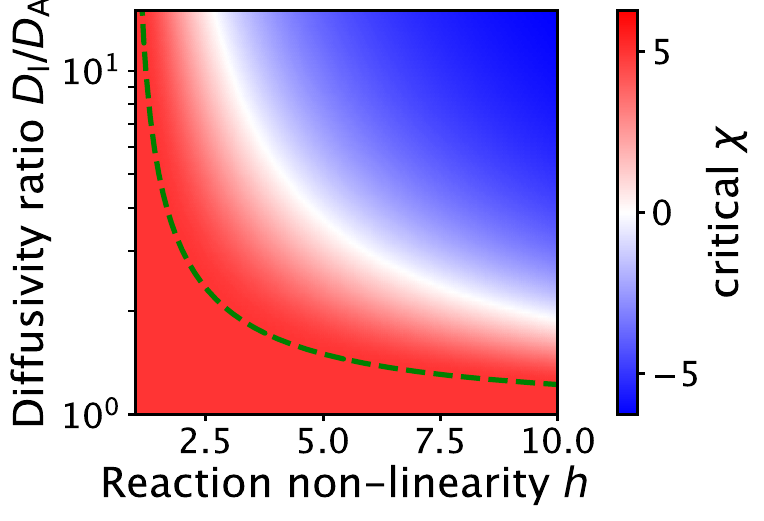}
  \caption{Critical $\chi$ as a function of the reaction non-linearity and the ratio of diffusivities at $\phi_0=0.2$. The green curve represents $D_\mathrm{I}/D_\mathrm{A}=(h+1)/(h-1)$ below which $\chi=1/\phi_0=5$. The values above the green curve are obtained using Eq.~(\ref{eq:S_chi_sol}). The white line corresponds to $\chi=0$.} 
  \label{fig:chi_h_DI} 
\end{figure}

In the opposing case of  $a>0$, the wave vector associated with the most unstable mode obeys $\diff (\det \Tjac )/\diff (\vect q^2)=0$ and thus reads $\vect q^2_{\text{max}}=-b/(2a)$, or,
\begin{align}
  \label{eq:S_qmax}
  \qMax^2
  =\frac{k}{4 D_A} \frac{1}{\eta}\left(\frac{ \eta (h-1)-(h+1)}{1-\phi_0 \chi  }-\frac{1+\eta}{1+2 \psi + \phi_0\chi }\right)
  \;.
\end{align}
If $-b/(2a)<0$, we have $\det(\Tjac)>\det (\tilde{\tens J}(0))$, implying that the homogeneous state is stable since $\det (\tilde{\tens J}(0))=k^2$.
However, if  $-b/(2a)>0$, 
the minimum value of $\det (\Tjac)$ is $\min_{\vect q} \det(\Tjac)=\det(\tilde{\tens J}(\vect q=\qMax))$.
In order to make the uniform state unstable, we thus require 
\begin{eqnarray}
  \label{eq:S_det_ineq}
  \det\left(\tilde{\tens J} \left(\vect q^2=-\frac{b}{2a}\right)\right)<0
  \;.
\end{eqnarray}
Consequently, the homogeneous state is unstable if
\begin{widetext}
\begin{align}
  \eta &>\frac{h+1}{h-1},
&
	h&>1,
& \text{and} &&
  \frac{ -(\eta -1)^2   (1+2\psi)h^2+2 \left(\eta ^2-1\right)  \left(1+\psi\right)h+8 u (\psi +1)-16 \eta  \psi}{\phi _0 \left(16 \eta +(\eta -1)^2 h^2\right)}&<\chi<\frac{1}{\phi_0}\;,
\end{align}
where $u=\sqrt{(\eta -1) \eta  (\eta  (h-1)+h+1)}$.
Taken together, we get three bifurcation curves in the parameter space that we are interested in:
\renewcommand{\theenumi}{(\roman{enumi})}%
\begin{enumerate}
\item $\chi=\chi^-_\mathrm{spin}=-1/(\phi_0(1-2\phi_0))$ for all $h>1$ and $\eta>1$. The uniform state is unstable for $\chi<\chi^-_\mathrm{spin}$.
\item  $\chi=\chi^+_\mathrm{spin}=1/\phi_0$ when $h<(\eta+1)/(\eta-1)$, or, equivalently, $\eta<(h+1)/(h-1)$. The uniform state is unstable for $\chi>\chi^+_\mathrm{spin}$.
\item In between these extreme cases, the critical interaction is given by
  \begin{align}
    \label{eq:S_chi_sol}
   \chi= \chi^*=\frac{ -(\eta -1)^2   (1+2\psi)h^2+2 \left(\eta ^2-1\right)  \left(1+\psi\right)h+8 u (\psi +1)-16 \eta  \psi}{\phi _0 \left(16 \eta +(\eta -1)^2 h^2\right)}\;,
  \end{align}
  when $\eta >(h+1)/(h-1)$ and $h>1$. The uniform solution is unstable when $\chi>\chi^*$.
\end{enumerate}
\end{widetext}
\figref{fig:h_chi_h_DI} summarizes these results and shows that the stability analysis of this limit of weak interactions (red lines) is very close to the full numerical stability analysis (black lines).
Furthermore, \figref{fig:chi_h_DI} shows the critical value $\chi^*$ as a function of $h$ and $\eta=D_I/D_A$.
For weak interactions (white region, $\chi^*\approx 0$), both decreasing $h$ and decreasing $D_I/D_A$ leads to higher value of $\chi^*$, which agrees with the special cases shown in Figs.~1 and 3 in the main text.
Another interesting observation is that the chemical reaction rate $k$ does not appear in \Eqref{eq:S_chi_sol}, implying the stability-instability transition is independent of the specific value of $k$, consistent with the results shown in \figref{fig:reaction} of the main text.

\subsection{Analysis of most unstable wave length}
We next use the approximate model in the limit of weak interactions to analyze the pattern length scales~$\ell$ shown in \figref{fig:reaction} of the main text and \figref{fig:overview_length_scale}. 
Here, $\ell = 2\pi/|\vect q_\mathrm{max}|$ is associated with the most unstable mode given by \Eqref{eq:S_qmax}.
In particular, we find
 \begin{eqnarray}
  \frac{\mathrm{d}q^2_{\mathrm{max}}}{\mathrm{d}h}=\frac{k}{4 D_A} \frac{1}{\eta}\frac{\eta-1}{1-\phi_0 \chi}>0
 \end{eqnarray}
 if $\eta>1$, implying that the length scale $\ell$ decreases with larger reaction non-linearity $h$.
 Similarly, we find
 \begin{multline}
  \frac{\mathrm{d}q^2_{\mathrm{max}}}{\mathrm{d}\eta}=\frac{k}{4 D_A} \frac{1}{\eta^2}\left(\frac{1+h}{1-\phi_0 \chi}+\frac{1}{1+2 \psi + \phi_0\chi }\right)>0
 \end{multline}
 if $\chi<1/\phi_0$, so $\ell$ decreases as the diffusivity ratio $\eta=D_I/D_A$ increases.
Moreover,
\begin{multline}
  \frac{\mathrm{d}q^2_{\mathrm{max}}}{\mathrm{d}\chi}=\frac{k}{4 D_A} \frac{\phi_0}{\eta}
  	\biggl(\frac{ \eta (h-1)-(h+1)}{(1-\phi_0 \chi)^2  }
\\
	+\frac{1+\eta}{(1+2 \psi + \phi_0\chi)^2 }\biggr)>0
 \end{multline}
if $\eta>(h+1)/(h-1)$.
Consequently, the length scale~$\ell$ also decreases as $\chi$ increases.
All these variations of $\ell$ capture the trends observed in \figref{fig:reaction} of the main text and \figref{fig:overview_length_scale}.
In the limit $\eta\rightarrow \infty$, we furthermore find
 \begin{eqnarray}
  q^2_{\mathrm{max}}=\frac{k}{4 D_A} \left(\frac{ h-1}{1-\phi_0 \chi  }-\frac{1}{1+2 \psi + \phi_0\chi }\right)
  \;,
 \end{eqnarray}
 implying the length scale~$\ell$ converges to a constant value for a given $k$ and $h$.
 In contrast, for given $k$ and $\eta$, we have
\begin{eqnarray}
  q^2_{\mathrm{max}}=h\frac{k}{4 D_A} \frac{1}{\eta}\left(\frac{ \eta -1}{1-\phi_0 \chi  }\right)
 \end{eqnarray}
for large $h$, so increasing the reaction non-linearity can in principle result in arbitrarily short~$\ell$.

\section{Phase separation in passive system}
\label{sec:passive_system}
To evaluate the influence of phase separation, we investigate the passive system without chemical reactions ($k=0$).
In this case, the stationary states of \Eqref{eqn:pde} in the main text are equivalent to the minima of the energy given in \Eqref{eqn:free_energy}.
The behavior of this system is characterized by the phase diagram given by the binodal and spinodal curve~\cite{weber2019physics, Zwicker2022a,mao2019phase}.

For associative phase separation ($\chi<0$), the activator~$A$ and the inhibitor~$I$ are perfectly co-localized, $\phi_A = \phi_I = \phi$.
Therefore, the non-dimensionalized free energy density of the system simplifies to 
\begin{eqnarray}
	f(\phi) = 2\phi\log(\phi)+(1-2\phi)\log(1-2\phi)+\chi\phi^2
	\;,
\end{eqnarray}
with associated chemical potential $\mu(\phi) = 2 \chi  \phi -2 \log(\phi^{-1}-2)$ and pressure $p(\phi) = \chi  \phi ^2-\log (1-2 \phi )$~\cite{weber2019physics}.
The point where the homogeneous state $\phi(\vect r)=\phi_0$ becomes unstable is given by $f''(\phi_0) = 0$, which results in the spinodal curve 
\begin{equation}
	\chi^-_{\mathrm{spin}}=-\frac{1}{\phi_0(1-2\phi_0)}
	\;.
\end{equation}
We determine the associated coexistence (binodal) curve assuming the system separates into two phases $\alpha$ and $\beta$ with $\phi_A^\alpha=\phi_I^\alpha=\phi^\alpha$ and $\phi_A^\beta=\phi_I^\beta=\phi^\beta$.
The coexistence conditions, $\mu(\phi^\alpha)=\mu(\phi^\beta)$ and $p(\phi^\alpha)=p(\phi^\beta)$, are then solved by
\begin{eqnarray}
	\chi^-_{\mathrm{bin}}= -\frac{8 \mathrm{arctanh}(1-4\phi_0)}{1 - 4\phi_0}
	\;.
\end{eqnarray}

For segregative phase separation ($\chi > 0$), the activator and the inhibitor repel each other.
Assuming the amount of $A$ and $I$ are equal and the interactions are symmetric, the solvent fraction is identical in both phases,  implying $\phi_A^\alpha=\phi_I^\beta$ and $\phi_A^\beta=\phi_I^\alpha$ as well as $\phi_A = \phi$ and $\phi_I = 2\phi_0 - \phi$.
The associated free energy density reads
\begin{multline}
	f(\phi)=\phi  \log (\phi )+(2 \phi_0-\phi) \log (2 \phi_0-\phi)
	\\+(1-2 \phi_0) \log (1-2 \phi_0)+\chi\phi(2 \phi_0-\phi)
	\;.
\end{multline}
Similar to the calculation above, we obtain the spinodal point $\chi^+_{\mathrm{spin}}$ and the binodal point $\chi^+_{\mathrm{bin}}$,
\begin{align}
	\chi^+_{\mathrm{spin}} &= \frac{\phi_0}{\phi(2 \phi_0 - \phi)}
	& \text{and} &&
	\chi^+_{\mathrm{bin}} &= \frac{\log \left(\frac{2 \phi_0}{\phi }-1\right)}{2 \phi_0-2 \phi }
	\;,
\end{align}
which now depend on $\phi$ and $\phi_0$.
To determine the minimal value of the interaction where (spontaneous) phase separation can be expected, we minimize both expressions with respect to $\phi$, which results in $\chi^+_{\mathrm{spin}} = \chi^+_{\mathrm{bin}} = 1/\phi_0$.

\begin{figure}[tb]
  \centering
  \includegraphics[width=\columnwidth]{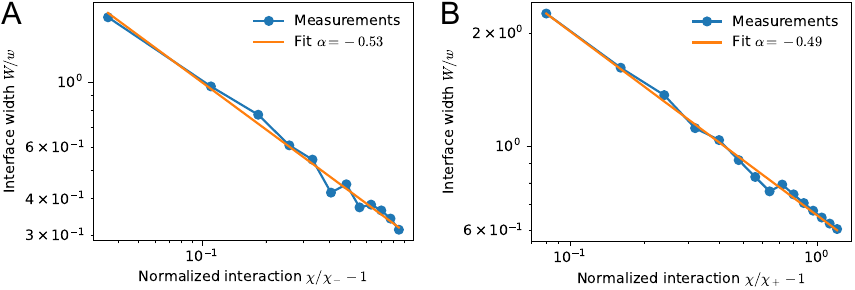}
  \caption{Width of the interface~$W$ in a phase separating system as a function of interaction strength~$\chi$ for $k=0$ and $\phi_0=0.2$.
  (A) Power law scaling of $W$ as a function of the normalized interaction strength for associative phase separation ($\chi < \chi_-$).
  (B) Power law scaling of $W$ as a function of the normalized interaction strength for segregative phase separation ($\chi > \chi_+$).
  (A--B) Interfacial widths where measured by fitting a hyperbolic tangent function to equilibrium states determined numerically on grids of length $60$ with $150$ support points.
  } 
  \label{fig:interfacial_width} 
\end{figure}

In the main text, we show binodal points for $\phi_0 = 0.2$, resulting in $\chi_- = \chi^-_\mathrm{bin} = -8.11$ and $\chi_+ = \chi^+_\mathrm{bin} = 5$.
\figref{fig:interfacial_width} demonstrates that the width of equilibrium interfaces is of the order of the microscopic length~$w$ and increases with proximity to the binodal lines given by $\chi_\pm$.
The scaling exponent $-\frac12$ is consistent with the behavior close to the critical point~\cite{Cahn1958}.

\end{appendix}
\clearpage

\bibliographystyle{apsrev4-2}
\bibliography{../bibli}

\end{document}